\RequirePackage{fix-cm}
\documentclass{svjour3}                     
\smartqed  
\usepackage{graphicx}
\usepackage[dvipsnames]{xcolor}

\begin{document}

\title{CovidSens: A Vision on Reliable Social Sensing for COVID-19}


\author{Md Tahmid Rashid \and Dong Wang}


\institute{Md Tahmid Rashid \at
              Department of Computer Science and Engineering \\
              Notre Dame, IN 46556, USA\\
              \email{mrashid@nd.edu}           
           \and
           Dong Wang \at
              Department of Computer Science and Engineering \\
              Notre Dame, IN 46556, USA\\
              \email{dwang5@nd.edu}
}

\date{Received: date / Accepted: date}

    \maketitle
    \begin{abstract}
With the spiraling pandemic of the Coronavirus Disease 2019 (COVID-19), it has becoming inherently important to disseminate accurate and timely information about the disease. Due to the ubiquity of Internet connectivity and smart devices, social sensing is emerging as a dynamic AI-driven sensing paradigm to extract real-time observations from online users. In this paper, we propose CovidSens, a vision of social sensing based risk alert systems to spontaneously obtain and analyze social data to infer the state of the COVID-19 propagation. CovidSens can actively help to keep the general public informed about the COVID-19 spread and identify risk-prone areas by inferring future propagation patterns. The CovidSens concept is motivated by three observations: 1) people have been actively sharing their state of health and experience of the COVID-19 via online social media, 2) official warning channels and news agencies are relatively slower than people reporting their observations and experiences about COVID-19 on social media, and 3) online users are frequently equipped with substantially capable mobile devices that are able to perform non-trivial on-device computation for data processing and analytics. We envision an unprecedented opportunity to leverage the posts generated by the ordinary people to build a real-time sensing and analytic system for gathering and circulating vital information of the COVID-19 propagation. Specifically, the vision of CovidSens attempts to answer the questions: How to distill reliable information about the COVID-19 with the coexistence of prevailing rumors and misinformation in the social media? How to inform the general public about the latest state of the spread timely and effectively, and alert them to remain prepared? How to leverage the computational power on the edge devices (e.g., smartphones, IoT devices, UAVs) to construct fully integrated edge-based social sensing platforms for rapid detection of the COVID-19 spread? In this vision paper, we discuss the roles of CovidSens and identify the potential challenges in developing reliable social sensing based risk alert systems. We envision that approaches originating from multiple disciplines (e.g., AI, estimation theory, machine learning, constrained optimization) can be effective in addressing the challenges. Finally, we outline a few research directions for future work in CovidSens.
\end{abstract}

\keywords{Social sensing \and COVID-19 \and coronavirus \and disease tracking \and real-time \and information distillation.}
    \section{Introduction} \label{sec:intro}
In this era of big data and pervasive Internet connectivity, social sensing is emerging as a dynamic AI-driven sensing paradigm that utilizes observations by humans and devices coupled with powerful AI devices (e.g., dedicated AI system-on-a-chip (SOC)) to obtain information about the physical world~\cite{ignatov2018ai,IPSN:12}. In this vision paper, we present CovidSens, the notion of real-time risk analysis and alerting systems based on social sensing to obtain the situational awareness and guide the intervention motives for Coronavirus Disease 2019 (COVID-19). According to the most recent statistics, there are more than 1.5 million confirmed cases of COVID-19 and above of 89,660 deaths spread across 50 states in US~\cite{cdc2020,point2020}. Most of the above cases happened within one week’s time (i.e., between March 29, 2020 and April 04, 2020) and the current trend seems to be ever increasing~\cite{point2020}. As the outbreak of COVID-19 progresses, circulating information about the spread in an accurate and timely manner has grown ever important. However, with heightening uncertainty and commotion among the general public, the communication of timely and accurate information to intended recipients is a challenging task. While official warning channels and news agencies have served an active role in informing the public about the spread, they often fall short in terms of pace. It is apparent that the official warning channels and news media take a while to confirm and disseminate the information regarding the outbreak of a new disease~\cite{vos2016social}. By contrast, information propagation across the social media and crowdsensing platforms is inherently faster than traditional news media~\cite{wang2019age}. For example, during the 2013 Boston Marathon Bombing, news about the first bomb explosion and the arrest of the suspect were posted in Twitter several minutes before news agencies made announcements~\cite{haddow2015social,haddow2013disaster}. After the onset of Cholera outbreak in Haiti in 2010, the knowledge regarding the outbreak was first obtained from social media, which occurred weeks before officials confirmed the case of the outbreak~\cite{chunara2012social}. Such cases exemplify the importance of social sensing during emergency scenarios such as now during the COVID-19 outbreak.

The CovidSens concept is thus motivated by three observations during this global crisis of COVID-19. First, people tend to actively convey their state of health and experience of the virus via online social media since the onset of the COVID-19. For instance, at one given day, 6.7M people talked about coronavirus on social media~\cite{forbesmillions}. Second, people report their observations on social media relatively faster than the official warning channels and news agencies that make formal announcements. As such, knowledge contribution and discovery through social sensing may offer more effective news transmission~\cite{wang2019age}. Third, online social media users, who report their observations of COVID-19, are frequently equipped with powerful mobile devices with rich processing capabilities~\cite{ignatov2018ai}. Such devices can execute complex AI models to distill information about the COVID-19 spread at the edge, potentially expediting the data analysis~\cite{zhang2019social}. Given these premises, we perceive an unprecedented opportunity to leverage the posts generated by the social media users to build a complete AI-driven analytics framework for rapidly gathering and circulating vital information of the COVID-19 propagation.

\begin{figure}[!htb]
 \vspace{-0.1in}
    \centering
    \includegraphics[width=8cm]{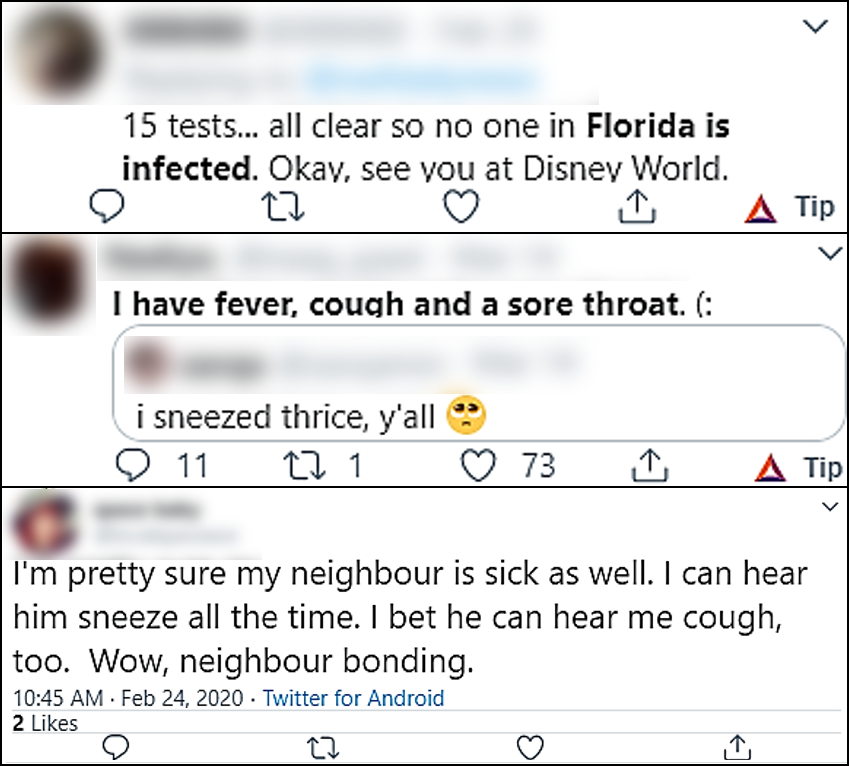}
    \caption{Tweets posted during the COVID-19 outbreak}
    \label{fig:realtweets}
\end{figure}

Let us consider a few tweets posted during the course of the COVID-19 spread across the US in Figure~\ref{fig:realtweets}. These tweets express the experiences and observations of individuals about the COVID-19. If such tweets could be analyzed using state-of-the-art AI algorithms to identify regions affected by COVID-19 and determine the rate of the spread, it might potentially expedite the alleviation of the adverse effects of the virus. In addition, by parsing the location and movement data from smartphones and social media posts to detect crowds or mass gatherings while respecting user privacy, government agencies and the mass public could be informed about the more risk-prone areas of a city during the COVID-19 outbreak~\cite{cnbcgoogle}. 
This could potentially help to divert people away from more crowded locations and hence reduce the spread of the disease.

While the CovidSens vision promises opportunities for a robust social sensing based information distillation and alert service for the COVID-19 spread, several technical challenges exist in the way of building such a system to autonomously gather and distribute real-time development of the disease to the general public. In contrast to traditional disaster response systems (e.g., for floods or forest fires), one unique goal of CovidSens is to obtain knowledge of the dynamics of the disease spread (e.g., inferring the stages of the disease among people). The first challenge is, therefore, to build a social sensing data collection platform that is able to spontaneously obtain the relevant social signals about symptoms, cases, and fatalities of COVID-19 from the online social media users. The second challenge lies in developing reliable data analysis models based on adaptive AI architectures~\cite{khan2019survey} that can extract credible information of the disease spread from the noisy, sparse, and unstructured social data contributed by unvetted human sources such as the tweets in Figure~\ref{fig:realtweets}. The third challenge exists in handing the huge volumes of social data about the COVID-19 outbreak that varies widely (e.g., across text, image, video, and audio data). The fourth challenge is how to distill information of the COVID-19 spread by customizing existing AI algorithms to run on the individually owned edge devices that are originally designed to run in a centralized fashion. The fifth challenge is to circulate the extracted information about the disease spread to the general public in a timely and efficient manner so that they can plan their actions accordingly. The sixth challenge lies in designing effective alert systems that consider the human aspect of the problem (i.e., handling people's reaction to alerts like fear, concern, or ignorance). The seventh challenge is combating the misinformation spread in the social media where people tend to report rumors or falsified facts of the COVID-19 spread. 

The CovidSens aims to overcome the above limitations by providing a more reliable and timely COVID-19 monitoring and alerting system for the mass population based on social sensing. We envision a dynamic and scalable AI-driven information retrieval and dispatching system for the general public based on data derived from multiple sources (e.g., social media, crowdsourced platforms, Unmanned Aerial Vehicle (UAV)) to quickly and effectively inform about the COVID-19 spread using a combination of smartphone applications, UAVs, message boards, or other modes of information dispersal. We expect this service to be important and useful for people who live in or travel to the affected areas, allowing them to take special precautions and be well prepared. The successful development of such systems can potentially help both authorities and general public respond more quickly and efficiently to COVID-19 and eventually help save more lives.

We acknowledge the potential to employ interdisciplinary techniques from deep learning, machine learning, estimation theory, game theory, online social media analysis, distributed systems, and mobile phone applications to develop effective CovidSens systems. Research along the realm of CovidSens is important because the COVID-19 is spreading rapidly in many countries worldwide and a timely alerting system that explores the rich real-time information streaming on social media is yet to be developed. The results of this research can pave the way for studying and tackling COVID-19 around the world. 

The rest of the paper is organized as follows. In Section 2, we discuss a few state-of-the-art works in the direction of CovidSens. In Section 3, we explore potential real-world applications of CovidSens. We identify the a few likely challenges in implementing a successful CovidSens system in Section 4. Afterwards in Section 5, we highlight a set of research directions for future work aligning with CovidSens to contain the COVID-19 spread. Finally, we conclude our vision of CovidSens in Section \ref{sec:conclusion}.
	\section{Related Works} \label{sec:related}

\subsection{Social Sensing}
Social sensing is rapidly progressing as a pervasive sensing paradigm where humans are used as sensors to attain situational awareness about the physical world~\cite{wang2019age}. Examples of social sensing applications include predicting poverty in developing countries~\cite{smith2013ubiquitous},  studying human mobility in urban areas~\cite{noulas2012tale}, identifying traffic abnormalities~\cite{zhang2020graphcast,wang2013credibility}, monitoring the air quality \cite{zhang2019online}, tracking social unrest~\cite{al2014crowd} and disasters~\cite{marshall2016mood,wang2013recursive},  and detecting wildfire~\cite{boulton2016using}. A comprehensive survey of social sensing schemes is provided in~\cite{wang2015social}. Zhang \textit{et al.} developed a scalable approach to obtain data veracity in social sensing~\cite{zhang2018scalable}. Xu \textit{et al.} developed a framework for semantic and spatial analysis of urban emergency events using social media data~\cite{xu2016participatory}. Zhang \textit{et al.} presented a constraint-aware truth discovery model to detect dynamically evolving truth in social sensing~\cite{zhang2017constraint}. More recently, there is an advent of social-media-driven drone sensing (SDS) approaches that address the data reliability issue of social sensing by integrating social signals with physical UAVs~\cite{rashid2020socialdrone}. While existing social sensing approaches aim to provide pervasive sensing, they are not tailored specifically to monitor the COVID-19 outbreak. Compared to traditional social sensing applications, CovidSens not only requires an inference of the data veracity, but also how the COVID-19 outbreak can progress across regions based on indications from social media posts (e.g., posts about crowded subways could indicate high risk of COVID-19 risk spread).  Thus, it remains a critical task to develop a reliable social sensing model that can accurately monitor the COVID-19 spread.

\subsection{Disease Outbreak Investigation}
In recent times, disease tracking based on epidemiological data has been an important avenue of research. Several studies have independently explored the feasibility of using social media and crowdsensing for detection, tracking, and analytics of contagious disease outbreaks~\cite{schmidt2012trending,charles2015using}. For example, Google launched a real-time influenza surveillance system, namely Google Flu Trends~\cite{wilson2009interpreting}, to monitor influenza spread by analyzing search terms related to illness symptoms. Kalogiros \textit{et al.} developed Allergymap, a crowdsensing-based disease identification system for allergen season onsets and allergy patient stratification~\cite{kalogiros2018allergymap}. Krieck \textit{et al.} studied the possibility of analyzing Twitter data for infectious disease surveillance~\cite{krieck2011new}. Chester \textit{et al.}~\cite{chester2011use} carried out bacterial outbreak investigation based on web forum posts about sick participants from a bike race. Despite the advances in disease monitoring techniques, current schemes have not been designed to handle the exponential progression of the COVID-19 pandemic and provide reliable risk alert in the context of CovidSens. Therefore, it entails a more rapid information distillation and processing system that can track the COVID-19 spread in real-time.

\subsection{Automated Disease Warning and Alert Systems}
While traditional health systems play an important role in alerting the general public about infectious diseases, their slow information progression have necessitated the adoption of automated warning and alert systems~\cite{schmidt2012trending}. Brownstein \textit{et al.} contributed a few early works in this domain by developing: i) a series of interactive websites, HealthMap and Flu Near You~\cite{schmidt2012trending,brownstein2008surveillance}, and ii) a smartphone application called Outbreaks Near Me~\cite{freifeld2008healthmap} to present vital information about outbreaks of various illnesses around the world. Toda \textit{et al.} explored the effectiveness of a text-messaging system for notification of disease outbreaks in Kenya~\cite{toda2016effectiveness}. Yu \textit{et al.} developed ProMED-mail, an early warning system for emerging diseases~\cite{yu2004promed}. Carter studied the possibility of a tweet-based information dispersal system to facilitate the containment of Ebola in Nigeria~\cite{carter2014twitter}. The above approaches are known to provide disease warning with reasonable effectiveness. However, it is an even more challenging task to develop a real-time COVID-19 spread indicator for CovidSens that uses both social-media and crowdsourced data, and also transmit the news of the spread to the general public in real-time.

\subsection{COVID-19 Spread Monitoring}
With the emergence of the COVID-19 outbreak, several streams of research have introduced methods to monitor the COVID-19 propagation. Sun \textit{et al.}~\cite{sun2020early} proposed the first study that harnesses crowdsourced data from several social media sources to monitor the COVID-19 spread. Schiffmann~\cite{schiffmann2020} developed an informative web portal that aggregates news from myriads of news sources to present latest information on COVID-19 spread. The Johns Hopkins Center for Systems Science and Engineering (JHU CSSE) developed an interactive online dashboard to track and present worldwide reported cases of COVID-19 in real-time~\cite{dong2020interactive}. An online community of international students and professionals, called 1point3acres, developed a web-based real-time COVID-19 news aggregator to track the state of the spread in the US and Canada~\cite{1point3acres}. A mobile app has been developed by the Singapore government to leverage crowdsourced information to locate community transmission of COVID-19~\cite{singapore2020}. A key drawback of the above tools is that they possess partial autonomy, requiring some degree of manual efforts to validate the information of the COVID-19 spread before presenting them online~\cite{schiffmann2020,1point3acres}. During this evolving COVID-19 outbreak, delays are undesirable. Therefore, a significant limitation exists in existing approaches to spontaneously track the COVID-19 propagation and disseminate the information to the end users.

\subsection{AI-driven Disease Prediction}
The growing demand for intelligent application domains like autonomous driving, robotics, computational medicine, computer vision, and natural language processing call for reliable AI-driven information distillation systems~\cite{abiodun2018state}. In the recent past, several studies have used AI for diagnosis, identification, and monitoring of infectious diseases using data collected from various sources (e.g., past disease records, social media posts, wearable sensors)~\cite{barrat2014measuring,kawtrakul2007framework,torres2016tracking}. Babu \textit{et al.} applied Grey Wolf optimization and recurrent neural networks (RNN) on patient symptom data for early disease detection and response~\cite{babu2018medical}. Du \textit{et al.} proposed a convolutional neural networks (CNN)-based approach for measles risk identification by analyzing public perception of measles outbreak from Twitter data~\cite{du2018public}. Torres \textit{et al.}~\cite{torres2016tracking} developed an artificial neural network (ANN)-based dengue tracking system based on prior infection data. Mahalakshmi \textit{et al.} built a Zika virus outbreak prediction system from symptom data based on multilayer perception (MLP) neural networks~\cite{mahalakshmiprediction}. However, despite the usefulness of existing approaches, due to the lack of sufficiently sized datasets with high quality labels on COVID-19, a key concern in AI-driven COVID-19 detection is ending up with underfitted and biased AI models that could yield erroneous prediction~\cite{naude2020artificial}. Moreover, while the above systems utilize efficient AI architectures for prediction of specific diseases, they have not been tailored to handle the massive scale of the rapidly progressing COVID-19 spread that has heightened to a global pandemic. It is therefore a challenging task to develop scalable and adaptive real-time AI-based  monitoring frameworks for COVID-19.

	\section{Real-world Applications} \label{sec:applications}
In this section, we highlight a few probable applications in real-world scenarios aligning with the CovidSens vision.

\subsection{Social-media-driven Disease Spread Indicator}
\begin{figure}[!htb]
 \vspace{-0.05in}
    \centering
    \includegraphics[width=8.0cm]{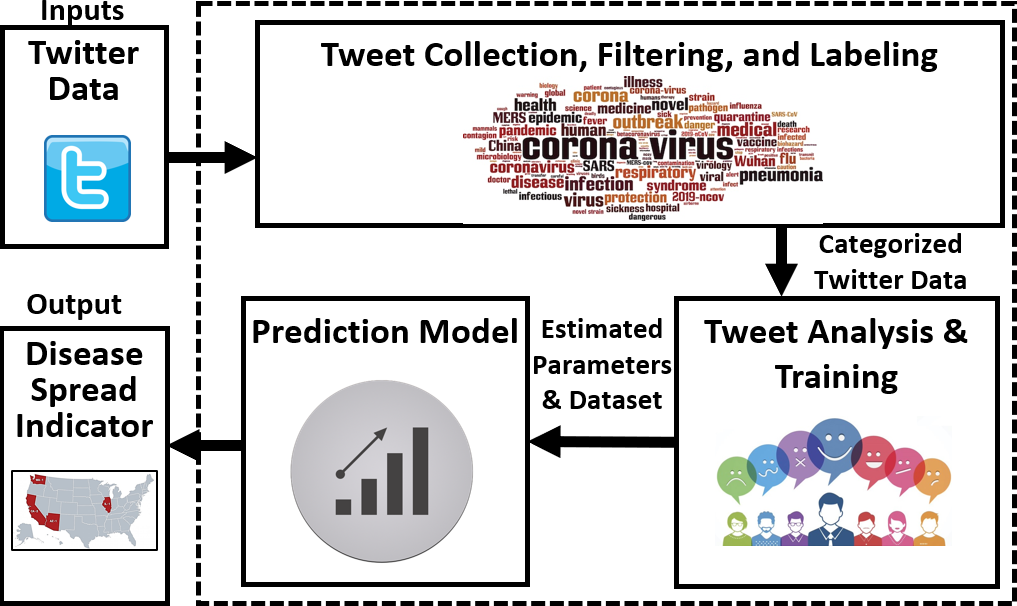}
    \caption{Overview of an SDSI system}
    \label{fig:block}
     \vspace{-0.05in}
\end{figure}

In a social-media-driven disease spread indicator (SDSI), social media posts related to COVID-19 are analyzed to attain the state of the spread~\cite{sun2020early}. An example of an SDSI architecture is illustrated in Figure~\ref{fig:block}. Initially, a real-time Twitter data crawler engine collects tweets indicating public opinions about the disease. The tweets are subsequently filtered and labelled into discrete categories based on the topics of discussions. A few examples of these topics can be: i) what regions are being frequently reported to be infected; ii) the time between people first talking about COVID-19 symptoms to deciding to be tested (i.e., how long the virus takes to show effect in people)~\cite{sun2020early}, iii) which age of people are expressing about symptoms the most; iv) how rapidly authorities are responding to the stimuli; and v) whether people are talking about other people they know getting recovered~\cite{sun2020early,cascella2020features}. Afterwards the labeled Twitter data are passed to a tweet analytics and training engine on a backend server. Specifically, the backend server will construct a clean and timely events summary about the COVID-19 spread by distilling relevant and reliable information from the massive amount of noisy, unstructured, and unvetted data feeds using adaptive AI algorithms such as Long Short Term Memory networks (LSTM) or Gated Recurrent Units (GRU)~\cite{ma2016detecting}. Lastly, a website or smartphone app will interact with end users to provide them warnings or alerts about the disease spread in their vicinity based on their queries. The analytics engine jointly analyzes the data veracity, source reliability, observation bias (e.g., under vs over estimation), as well as the likelihood of large-scale havoc launched by malicious users on social media using novel estimation theoretic, machine learning, and deep learning techniques. 

\subsection{Crowdsensing-based Disease Tracking}
\begin{figure}[!htb]
    \centering
    \includegraphics[width=7.5cm]{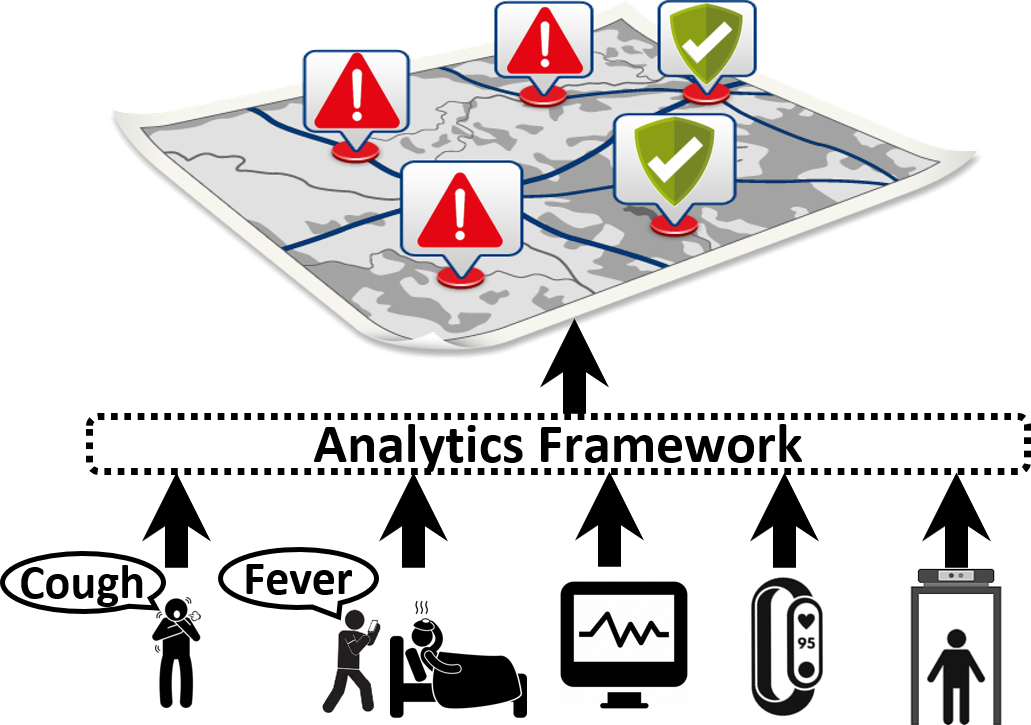}
    \caption{Overview of a CDT system}
    \label{fig:block2}
\end{figure}

Crowdsensing-based disease tracking (CDT) involves sensor networks and groups of people, with mobile devices capable of sensing, collectively sharing disease related information (e.g., early symptoms, nearby infected persons, deciding to self-quarantine)~\cite{sun2020early,haddawy2015situation}. CDT is fueled by the observation that individuals tend to proactively volunteer in contributing data about the COVID-19 spread using their smartphones, wearables, or other devices with sensors and connectivity~\cite{sun2020early}. In contrast to SDSI, CDT is relatively less pervasive and requires active participation of people and physical sensors. However, in return the data is less noisy and is hence more reliable. Figure~\ref{fig:block2} shows an example of a representative CDT system. A CDT may typically incorporate three main components. The first component is a data collection platform consisting of a network of users with a custom smartphone application to log data and a set of internet-of-things (IoT) devices (e.g., smart heart-rate monitors, activity trackers, thermal scanners). The smartphone application interacts with users and allows them to actively contribute their reports on the COVID-19 if they are willing to. If the users choose to input data, the app lets the users configure at what granularity (e.g., state, county, street, or N/A) they feel comfortable to share their location information. The second component is an analytics framework that applies relevant statistical analysis and AI techniques on the obtained data to infer probable regions of infection and safe zones~\cite{freifeld2008healthmap,haddawy2015situation}. To conserve bandwidth and expedite processing, the computational power of the smartphones can be harnessed to execute the AI algorithms at the edge. The third component is a smartphone application on the end users' mobile phones to visually represent the analyzed geospatial distribution of the inferred regions~\cite{freifeld2008healthmap}. The app can obtain the needed information from the backend server based on the users’ queries (e.g., checking the risk level of a particular area of interest) \cite{zhang2018towards}. In most cases the data collection, processing, and representation is carried out in the same smartphone application~\cite{freifeld2008healthmap}. Sun \textit{et al.} proposed one of the earliest crowdsourcing based COVID-19 outbreak detection system~\cite{sun2020early}. The Singapore and South Korea governments have launched mobile apps that utilize crowdsourced data to trace community transmission of the COVID-19~\cite{singapore2020}.

\subsection{UAV-based Health Surveillance and Alerting}
\begin{figure}[!htb]
        \centering
        \includegraphics[width=8.5cm]{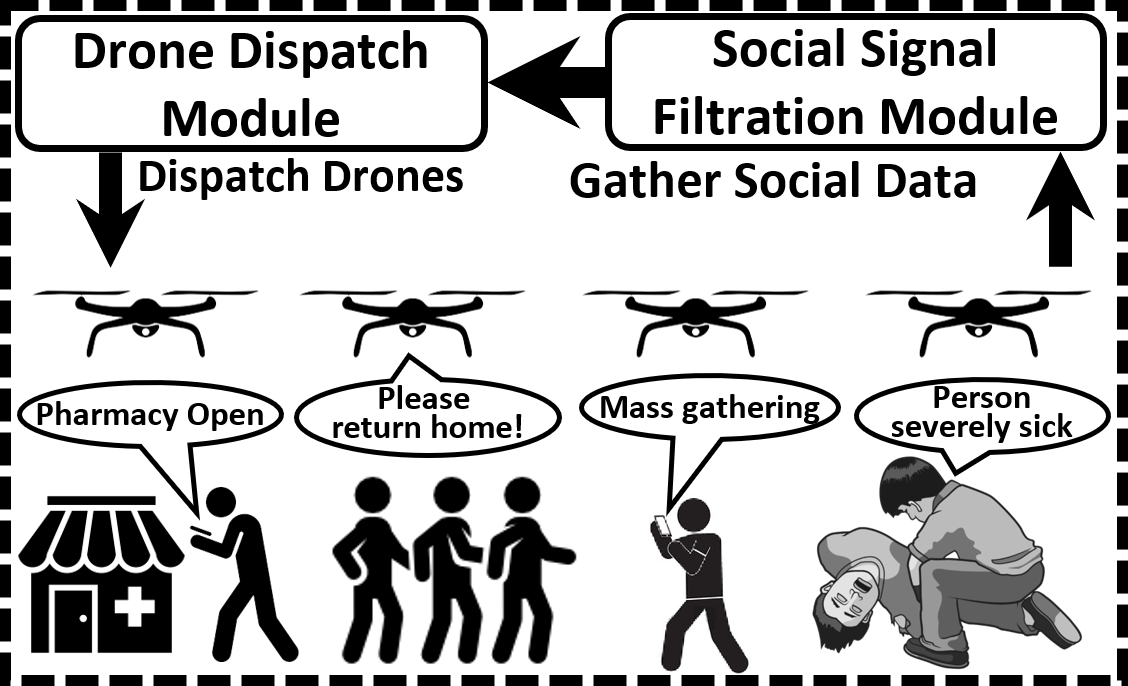}
        \caption{Overview of an UHSA system}
        \label{fig:block3}
    \end{figure}

The urgency of the COVID-19 outbreak has necessitated new dimensions for UAV-based health surveillance and alerting (UHSA) systems~\cite{minaeian2015vision}. With the help of onboard sensors (e.g., cameras, microphones), UAVs are able to gather intelligence remotely during a disease pandemic scenario where human patrol teams and ground units cannot operate due to risks of getting infected. For instance, UAVs can assist in detecting unwanted crowds of people along locked down areas of a city~\cite{minaeian2015vision}. Figure~\ref{fig:block3} demonstrates a representative UHSA model for mitigating the COVID-19 spread. The UHSA system responds to emergency requests by individuals through social media posts about unnecessary mass gatherings. Afterwards, the data is gathered in a backend server and processed using  social sensing approaches based on statistical analysis, deep learning, and machine learning for analyzing the truthfulness of the data. The information is then updated across nearby regions by raising verbal alerts through speakers installed on the UAVs. UAVs are also dispatched out to different areas of a city to spontaneously scan and obtain situational awareness about the region. Using the onboard sensors and image classification algorithms like Convolutional Neural Networks (CNNs), UHSA detects if people are breaking the rules during the lock down situation (e.g., by roaming outside, gathering in crowds). The framework may also locate and verify the availability of critical supplies using the UAVs (e.g., open pharmacy, grocery stores) based on the social media posts. Using the onboard speakers of the UAVs, the people breaking the rules are alerted to return home. One real-world example of UHSA during the COVID-19 ordeal is in California, USA where the law enforcement officials have resorted to utilizing drones for patrolling the state of California during the ongoing lock-down situation~\cite{caldrone2020}. During the COVID-19 crisis in China, UAVs have served multiple roles including post-epidemic aerial evaluation, alerting, and relief distribution to affected regions~\cite{ruiz2020uses}.

	\section{Research Challenges and Opportunities}
In this section, we present a set of prevalent research challenges and opportunities in the development of an effective CovidSens framework.

\subsection{Data Collection Challenge}
During the onset of rampant disease outbreaks like COVID-19, the primary objective of a CovidSens system is to collect information from the general public. However, several difficulties prevail to locate and obtain the relevant posts related to the COVID-19 spread. For instance, while conducting simple keyword based searches on obtained social media data, the desired keywords may indicate various other unwanted things (e.g., while the term ``sick'' is generally used to indicate people who are not doing well, it may also be used to express sarcasm by certain people). Several recent studies focused on mitigating this issue of data discovery by replacing simple keyword based searches with singular value decomposition (SVD) driven K-means clustering~\cite{nur2015combination}, adaptive sampling \cite{zhang2018light}, and recurrent neural network (RNN) based textual labeling process~\cite{jagannatha2016structured}. However, such methods still lag behind human perception in terms of accurately scanning for relevant input data. Thus, obtaining a collection of relevant social media data that directs to the right set of information remains an arduous task. Moreover, a great portion of social media data may eventually turn out to be redundant (e.g., retweets) or simply rephrased from a single original post~\cite{zanzotto2011linguistic}. On top of that, a good amount of social media data is observed to be transient and perishable \cite{zhang2019deeprisk}. For example, people may delete their previous posts and online repositories (i.e., Twitter and Facebook servers) hosting the posts may take them down for undisclosed reasons. In addition to that, social media APIs such as Twitter often impose various rate limitations which can heavily impede the data collection during disease outbreaks~\cite{makice2009twitter}. The data collection process for COVID-19 therefore necessitates a tool that can locate, obtain, and store the relevant information from users in real-time across social media channels.


\subsection{Data Reliability Challenge}
The concept of CovidSens is centered around the noisy and unreliable data generated by the unknown human sources on the social media~\cite{wang2013exploitation,wang2014maximum,wang2014provenance,wang2014surrogate}. One important task while harnessing social media for CovidSens is to extract trustworthy information from unreliable human sources with unknown source reliability~\cite{IPSN:12}. We define this as the \emph{data reliability} challenge in social sensing. Several truth discovery solutions have been developed to mitigate the data reliability problem. For instance, Wang et al. presented a framework to jointly estimate the reliability of data sources and the correctness of the reported measurements in social media posts using approaches from estimation theory~\cite{IPSN:12,wang2014using}. Zhang et al. built upon the previous framework to address the scalability and physical constraint challenges and employed the improved schemes to real-world social sensing applications~\cite{zhang2018scalable,zhang2017constraint}. Yin et al. developed Truth Finder, a probabilistic algorithm using iterative weight updates to improve the quality of the data in social sensing~\cite{yin2008truth}. While great efforts have been made on developing reliable social sensing solutions, certain limitations hinder these solutions from being applied in CovidSens to track COVID-19. 
One drawback of traditional social sensing schemes is that they solely rely on the noisy social media data and there no external means of validating the credibility of the input data during the COVID-19 epidemic~\cite{zhang2017constraint}. Existing methods are also not tailored towards disease outbreak detection, which may lead to prediction of false cases of COVID-19. For example, a person simply posting a symptom of breathing difficulty may not necessarily suffer from COVID-19. It may be required to analyze other traits of the patient based on earlier posts. 
Hence, it remains an unresolved challenge in CovidSens to develop reliable social sensing models that can explore the uncertainty in the input data and extract reliable signals. 

\subsection{Data Modality Challenge}
While data collection is an intrinsic challenge in using social sensing for tracking the COVID-19 spread, a greater difficulty exists in processing the rapidly generated incoming signals consisting of multitudes of features or dimensions~\cite{wang2015social}. This challenge is identified as \textit{data modality} in social sensing where large amounts of unfiltered and unstructured data with multiple modalities need to be processed~\cite{chu2016data,zhang2019transland,zhang2020trans,shang2019towards}. Specifically, data modality refers to the different variety or types of data prevalent in the social media such as text, image, location, audio, and video~\cite{birke2014big}. Moreover, each type can further encompass different dimensionality as well which makes the data modality challenge even harder. Examples of dimensionality in CovidSens can range along reports of: i) proximity to infected locations, ii) number of suspected cases, iii) number and types of symptoms, iv) intensity of symptoms (i.e., mild, moderate, or severe), v) recovery rate, vi) death rate, and vii) number of self-quarantined cases. Recent social sensing tools primarily focus on analyzing the text data in social media~\cite{zhang2018opinion}. This trend is advocated by the fact that image data processing involves heavy computation requirement~\cite{zhang2010image}. 

Consequently, existing methods do not focus on fusing multiple types of data which may potentially generate richer detection of COVID-19 propagation. For example, a person may tweet about having COVID-19, but based on an image posted with the tweet it may turn out that the person's symptoms have actually resulted from an allergic reaction instead~\cite{allergy}. Fusing text with other data such as image and location data may potentially yield more accurate prediction of the COVID-19 spread. Therefore, given the sheer volumes of multi-modal data generated by the social media users about the COVID-19 outbreak, solutions need to be developed to efficiently utilize the different modality of data. Moreover, since multi-modal data processing intrinsically demands greater compute power, care must be given to efficiently strike a trade-off between detection accuracy and computational complexity. A set of unsolved questions springing from the data modality challenge in CovidSens are: i) How to efficiently fuse the different types of social media data related to COVID-19 into one unified data stream? ii) How to design algorithms to process a wide variety of social data in real-time for an accurate prediction of the COVID-19 spread? iii) How to speed up the analysis of multi-modal data for faster COVID-19 spread detection by distributing the computation across multiple devices?

\subsection{AI-model Scalability Challenge}
Due to the global scale of the COVID-19 outbreak, it is important to resort to adaptive AI-based methods that can effectively monitor the state of the spread from the social sensing data across any region of the world in real-time. This necessitates the scalable AI algorithms that can be readily deployed across the edge devices (e.g., smartphones, IoT devices, drones) in order to reduce latency and bandwidth consumption, and yield faster information extraction for the COVID-19 spread. Unfortunately, existing AI schemes such as DNNs, MLPs, and RNNs have been originally developed for powerful centralized hardware (e.g., GPU clusters) and are not tailored for resource constrained smart devices residing at the edge of the network~\cite{li2018learning,zhangedgebatch,zhang2019heteroedge}. In particular, current AI algorithms are associated with model update processes that operate in a centralized fashion, which imposes a high network bandwidth requirement. In addition to that, mainstream AI models require extensive training to update the model parameters before being able to generate reliable predictions. Thus, even if the current AI algorithms could be improvised to run on the edge devices, due to their heavy computation requirements for the model training processes, they would drain the batteries of the portable edge devices faster~\cite{vance2019towards,zhang2018real,zhang2018cooperative}. A few open questions in CovidSens originating from the AI-model scalability challenge are: i) how to parallelize the AI model training process across the edge devices to speed up the model training and conserve network bandwidth? ii) How to optimize the AI algorithms to run efficiently on the energy constrained edge hardware? iii) How to modularize the AI algorithms so that they can be seamlessly deployed across a large number of edge devices without a single point of failure?

\subsection{Location Data Scarcity Challenge}
One recurring issue in social sensing is the user privacy whereby the personal information of the online users remains at risk of falling into the wrong hands~\cite{vance2018privacy}. Geo-location data shared by users can also be used to expose other private information as well (e.g., ethnicity, race, financial status) which social media users do not typically consent to share and are also not required by CovidSens applications. Thus, it has been observed that due to the concern of one's location and private information being exposed, many social media users tend to not share their location information while reporting their observations in the social media~\cite{zhang2018risksens,zhang2019riskcast,zhang2019syntax}. For example, in an independent study involving data collection for disaster related tweets, it was found that less than 10\% of the tweets were actually geo-tagged (i.e., contained geographical location of the users). As such, CovidSens applications that heavily rely on the location metadata from the social media posts to provide inference of the COVID-19 spread may under-perform when the number of geo-tagged social media are scarce. Recent literature has explored methods to work around this issue by exploiting spatiotemporal social constraints for location inference from social media posts~\cite{huang2017you}. However, such uni-dimensional approaches that rely on solely on the content of the social media posts may result in high estimation errors for the inferred locations. In order to precisely track the progress of the COVID-19 propagation, it is imperative to obtain the exact locations of the surges. Consequently, it is a challenge in CovidSens applications to design a solution that can mitigate the data scarcity issue which may eventually yield better sensing results for tracking the COVID-19 spread.

\subsection{Timely Presentation Challenge}
With the rapidly evolving circumstances during the COVID-19 outbreak, it is critical to present the information of the disease spread to the end users in a timely manner. This necessitates an information presentation system that can both process as well as present data of the disease propagation in real-time and keep people alerted. In the recent past, several methods have been implemented to present disease outbreak updates to the mass through means of interactive websites~\cite{schmidt2012trending,brownstein2008surveillance}. However, such methods of information distribution and collection solely rely on aggregating knowledge from different news portals and information websites which can lead to potential delays in alerting people about the most recent situation~\cite{wang2019age}. Due to their structured nature of information crawling and collating, existing web-based techniques cannot be directly applied to social sensing which encompasses unstructured and noisy social data~\cite{wang2019social}. In addition to that, websites and smartphone applications rely on the constant availability of both the Internet and a smart device, either of which may not be available in all circumstances. Thus, vital information may not reach all sectors of the population, especially with the elderly and less tech savvy individuals without access to computers and smart devices. Based on these grounds, it remains an open question in CovidSens on how to develop a reliable yet efficient mechanism that can rapidly deliver important messages and information regarding the COVID-19 spread to all segments of the population.

\subsection{Human Factor Challenge}
One important aspect to consider while dealing with social signals in CovidSens is the human component. Given the intensifying concerns and panic among the general public during the COVID-19, we acknowledge that people can be overly emotional, sensational, or biased in expressing their opinions in the social media or the crowdsensing applications~\cite{kim2016garbage}. Such behavior can potentially trigger misrepresented or misinterpreted observations and thus yield erroneous disease tracking results. Based on the above concerns, one critical challenge stemming from the human aspect of social sensing can be on deciding how to handle the mood of the population while containing the public concern at desirable levels. Moreover, it is imperative to study the human component closely and model how people react to the information presented to them through the warning and alert systems in CovidSens. Some individuals may turn out to be excessively sensitive and thus care must be taken so as not to develop the grounds for unnecessary panic or civil unrest. For example, during the Ebola epidemic in Liberia in 2014, riots broke out among the residents when officials raised alarms of the outbreak~\cite{fisman2014early}. On the other extreme of the spectrum, we also acknowledge that a certain proportion of the population have a tendency to be oblivious of the circumstances, neglect warnings, and remain excessively calm during this outbreak situation. The challenge of CovidSens is to strike a smooth balance between raising attention and providing assurance: at one end we need to calm people down while informing them of the situation but at the same time we also need to send out the message to remain well-prepared.

\subsection{Misinformation Spread Challenge}
\begin{figure}[!htb]
    \centering
    \includegraphics[width=8.5cm]{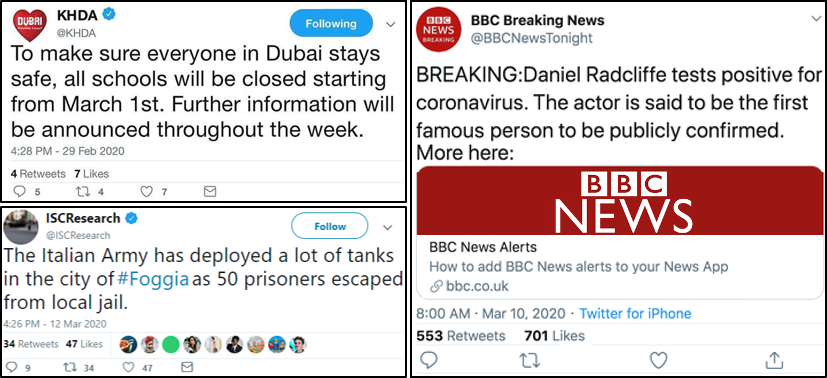}
    \caption{Tweets indicating fake news}
    \label{fig:faketweets}
\end{figure}

With the heightening concern of the COVID-19 spread, just as social media has served as a platform for attaining information, it has also served as the venue for sprouting misinformation. Due to the increased adoption of social sensing as a news source, misinformation spread on social media has remained an inevitable issue~\cite{yin2008truth}. This has caused social media giants such as Facebook and Google to conduct worldwide campaigns to fight the propagation of fake news~\cite{wingfield2016google}. Figure \ref{fig:faketweets} illustrates a collection of tweets referring to misinformation during the COVID-19 outbreak. The World Health Organization (WHO) has been forced to reallocate considerable resources to combat swathes of misinformation like these, which may potentially hinder COVID-19 monitoring efforts~\cite{who2020}. This phenomenon has been classified by WHO as an `infodemic’~\cite{who2020}. Social sensing tools, otherwise known as \textit{truth discovery} algorithms, are known to under-perform in the presence of widespread misinformation, which is common during disease outbreak scenarios. One obvious measure to address this issue is to acquire ground truth for validating the source reliability and event correctness. However, obtaining such ground truth is delay prone since it requires a significant amount of manual effort, but most importantly it is impractical during the course of virus breakouts where people should restrict locomotion and contact with other people. Therefore, it remains a critical challenge in CovidSens to construct an effective mechanism that can identify and isolate the misinformation spread to generate trustworthy social signals indicating the COVID-19 spread. 

	\section{Road-map for Future Work}\label{sec:roadmap}
In this section, we discuss a few potential directions for future work in the realm of CovidSens.

\subsection{Uncertainty Quantification in CovidSens}
We note that CovidSens relies on noisy and uncertain social-sensing data generated by unvetted data sources to monitor the COVID-19 spread. Thus, one domain for future work can be to mitigate the data reliability challenge for CovidSens applications. We observe that existing social-sensing tools or \textit{truth discovery} algorithms mainly prioritize on the data veracity or source reliability from the social media data. However, in a social-media-driven COVID-19 spread indicator application, the estimation confidence of a reported event's veracity is also crucial~\cite{wang2019social}. Consequently, it is important to determine the confidence level with which the COVID-19 propagation is predicted. For example, an inferred age demography with a low estimation confidence can easily lead to an erroneous conclusion on which age of people are most likely to be affected by COVID-19. In particular, further research can focus on rigorously quantifying the uncertainty of output results to evaluate and enhance the performance of the truth discovery algorithms. While the uncertainty quantification is well-studied in statistics and estimation theory, it is mostly overlooked in existing social sensing solutions since the performance of truth discovery algorithms are hard to inspect and humans are more likely to generate the claims with different degrees of uncertainty (e.g., affirmative assertions versus pure guesses)~\cite{wang2015confidence}. Based on this, one probable research direction is to develop a method to determine the confidence levels of detection by quantifying the uncertainty of the results in CovidSens applications.

Current literature on statistical analysis discusses principled approaches based on estimation theory. A set examples of techniques to quantify the uncertainty of the estimation results of the truth discovery algorithms are maximum likelihood estimation (MLE) and Cramer-Rao lower bounds (CRLB)~\cite{wang2013credibility,wang2011quantifying,wang2011bayesian,wang2012scalability}. While these methods have been tested to operate optimally to provide the desired uncertainty quantification, it still remains a critical challenge to formulate the truth discovery problems in CovidSens in a mathematically tractable way that would allow the uncertainty estimation tools to be applied upon. We envision that theories from multiple disciplines would be leveraged to cater to the uncertainty quantification problem in the CovidSens applications.

\subsection{Rumor Suppression and Fake News Detection}
One direction for future work for CovidSens is to combat the misinformation propagation challenge. Therefore, rumor suppression and fake news detection are indispensable for COVID-19 related misinformation spread containment. We acknowledge that rumors and misinformation in social media originates from the behaviour of individuals sharing what others share~\cite{wang2014provenance,kumar2014detecting}. Thus, it is beyond the scope of machine intelligence alone to contain the spread of rumors and misinformation entirely. Based on these premises, a few potential research questions can be: i) how to develop techniques that incorporate human intelligence along with machine intelligence to more accurately identify the rumors from true information about the COVID-19 spread? ii) How to investigate and identify the origin behind misinformation sharing from the social media posts? iii) How different demography (e.g., age groups, gender classes) react to misinformation about COVID-19 spread and how to utilize this knowledge to combat the misinformation propagation?

Several existing literature has proposed different fact determining techniques for analyzing and detecting falsified claims and rumors on social media using: i) Bayesian-based heuristic algorithms~\cite{yin2008truth}, ii) analyzing textual evidence with associated images~\cite{zhang2018fauxbuster}, and iii) considering physical constraints and temporal dependencies of the evolving truth~\cite{zhang2017constraint}. One new domain of research focuses on unifying the collective strengths of human intelligence (HI) and artificial intelligence (AI) to screen out misinformation in the social media~\cite{zhang2019crowdlearn}. Such approaches utilize HI-based crowdsourcing platforms such as Amazon Mechanical Turk (MTurk) in combination with existing deep neural networks (DNNs) and machine learning techniques, and can be used to classify social media posts about COVID-19 as veracious or falsified~\cite{zhang2019crowdlearn}.

\subsection{Mesh Network for News Aggregation and Circulation}
A stream of potential research can focus around mitigating the data collection and timely presentation challenges in CovidSens applications. In order to obtain information, traditional news media (e.g., CNN, BBC) rely on dedicated news reporters while social news aggregators (e.g., Digg, Reddit) rely on active voluntary participation of committed individuals \cite{shang2019vulnercheck}. A key drawback of such news collection approaches is that they entrust a central authority (i.e., a news agency or web administrator) to analyze and verify disease outbreaks like COVID-19, which may induce delays in deriving the COVID-19 propagation~\cite{wang2019age}. In contrast, a decentralized social-sensing based news aggregation and subscription service can potentially accelerate the news collection as well as distribution of information during the global pandemic of COVID-19~\cite{hong2012online}. A survey shows that 37\% of Internet users promulgated news content through social media posts on Facebook and Twitter~\cite{hong2012online}. With the proliferation of smart devices and people's tendency to post about being tested positive for COVID-19~\cite{testpositive,testpositive2,testpositive3} as well being tested positive on antibodies~\cite{antibody}, information about probable COVID-19 cases can propagate very fast through the social media. However, as identified earlier, a key hurdle is to develop a system that can spontaneously locate, obtain, and store the data from the social media platforms. Furthermore, after the COVID-19 related information is assembled, a system needs to be developed that can convey the processed information to the mass public. A set of important research questions are: i) how to efficiently filter and organize information contributed by diversified and unreliable sources? ii) How to compile the gathered information to an acceptable degree that each subscriber feels complacent in reading and trusting? iii) How to present the information to less tech savvy individuals with limited knowledge on computers and smartphones? iv) How to sustain the news aggregation and circulation during an Internet downtime?

A possible approach of information collection is to develop a real-time social media data collection and storage engine, such as Apollo~\cite{apollo2020}. One other potentially effective technique for information aggregation is to develop a dedicated crowdsensing-based smartphone application which allows users to readily report about COVID-19 related observations~\cite{freifeld2008healthmap}. Subsequently, a decentralized mesh network based news subscription service can be constructed from the collected data in the mobile app that is able to operate autonomously without a central authority. The service can be used to leverage the rich set of real-time observations of COVID-19 contained in the social data to explore the collective wisdom of common individuals without relying on dedicated news reporters. The entire service may be implemented within the aforementioned mobile app that can both collect the information of the COVID-19 spread from the online users and also present the prepared news to others~\cite{freifeld2008healthmap}. This process can virtually eliminate the existence of a central authority, hence reducing delays in information gathering and distribution in a CovidSens application. 

\subsection{Privacy Aware Location Discovery Based on Contextual Analysis}
CovidSens applications are inherently location data driven and hence a potential domain of research in CovidSens can be to address the location data scarcity challenge from the social media data. Specifically, studies can focus on determining the location of the COVID-19 related report origination points in the absence of the geo-location metadata in the posts. We emphasize that during inferring the event report locations from the social media data, care must be given to respect individual privacy from the system perspective, which if done improperly may lead to serious privacy breaches. For example, while a user's location information may be deduced from the text data in social media, it may also be used to infer other sensitive information such as job, ethnicity, race, financial status \cite{zhang2017large,zhang2019sparse}. Leakage of these information may place users at risk and lead to loss of confidence in the developed system~\cite{vance2018privacy}. Therefore, one important area of research in CovidSens can focus on how to develop privacy-aware location inference tools based on the contextual analysis of social media data that protects the identity and privacy of the users.

Once the user privacy is ensured, a good amount of opportunity exists in designing techniques to leverage the contextual information that is embedded within the text content of a social media post (toponym resolution). Moreover, images contained with posts can also be useful in extrapolating an accurate estimate of the social media report's origination sites~\cite{gallagher2009geo}. For example, an individual tweeting about COVID-19 symptoms claiming to be from a particular location can be given greater credibility if he or she posts with the image of the place. Another way to obtain the geo-location information of social media data can be to use image-based geocoding where subjects in the background of a posted image are cross-referenced with known landmarks or popular sites to find the location of the image~\cite{lin2010joint}.

People who post about disease symptoms in social media and ``follow" other social media users with similar symptoms may be co-located~\cite{gu2012fusing}. Intuitively, if one user's location can be determined, the location of the related users may be discovered as well. However, individuals may also reside very far from one another. For instance, two friends showing COVID-19 related symptoms may be located in two different cities. Thus, additional features from the social media data may be analyzed to infer other evidence for being co-located. Rich privacy-aware location inference schemes can be developed that fuse friend-follower networks with the contextual information embedded within texts in tweets to determine the whereabouts of COVID-19 spread~\cite{huang2017you,gu2012fusing}. An ensemble of solutions employing natural language processing (NLP)~\cite{dhavase2014location}, deep neural networks (DNNs), and social network analysis can be built to accurately infer the location information from the social media data~\cite{zhang2019crowdlearn,gallagher2009geo}.

\subsection{Edge Intelligence with Federated Learning}
One prospective domain for future research in CovidSens can be focused on addressing the scalability challenge of the AI models to effectively monitor the COVID-19 propagation from the social sensing data. In order to ensure that the most up-to-date information of the COVID-19 spread is available at any instant across any location, a large scale deployment of CovidSens is crucial. However, since traditional AI models are inherently built with a design philosophy that endorses centralized training \cite{zhang2019integrated}, they may not be a viable approach for such a global scale implementation of CovidSens. Therefore, in order to reliably analyze the obtained data related to COVID-19 across a global extent, we envision expandable AI architectures that can be spontaneously deployed across a massive number of edge devices.

With the growth of powerful edge devices (e.g., smartphones, IoT devices) and the demand for distributed model training over a large number of computing nodes, federated learning (FL) is gaining traction as a distributed AI training paradigm~\cite{konevcny2016federated}. In FL, a shared global AI model is trained from a collection of edge devices owned by end users, while retaining the training data within the edge devices~\cite{wang2019social}. By not transmitting the private data to a central server, FL manages to preserve user privacy and therefore foster trust among the participating users. The principle of FL aligns appropriately with our vision of scalable social sensing systems by shifting AI from the cloud to edge devices. However, there are still several open challenges in FL that need to be addressed before establishing effective CovidSens systems. One recurring issue in FL is the inconsistent availability of the edge devices, otherwise known as \textit{churn}~\cite{vance2019towards}. FL heavily relies on the participation of the edge devices for the training phase, which requires multiple iterations to converge to a global optima. Edge devices are owned by rational individuals who might abruptly leave in the middle of an ongoing AI model training process~\cite{wang2019social}. Moreover, edge devices might periodically evict tasks for power savings, or have a higher priority task to supplant the model training task. This could potentially negate the learning process, yielding poor model parameter training~\cite{vance2019towards}. Another limitation of many existing FL schemes is that they rely on synchronous model update operations~\cite{chen2019communication}. At every iteration of the model training, the server aggregates the model weights after receiving updates from all the clients. Due to the heterogeneity of the edge devices and the instability of network connections, all the devices cannot be guaranteed to have the same update interval~\cite{zhangedgebatch}. Thus, the server is prone to substantial downtime while needing to wait for all local updates before aggregation. In a CovidSens application, where time is a crucial factor, such delays are undesirable as they may slow the real-time prediction of the COVID-19 spread. Therefore, it is an open challenge to simultaneously handle the churn issue and develop asynchronous model training in FL for scalable CovidSens applications.

\subsection{Integration of Social Sensing with Physical Sensing}
As identified earlier, one key goal for developing effective CovidSens applications is to address the data reliability challenge stemming from the unreliable social media users. Beside uncertainty quantification, a strand of research to combat the data reliability challenge in CovidSens is to integrate social sensing with physical sensing paradigms (e.g., unmanned aerial vehicles (UAVs) and vehicular sensor networks (VSNs)) to verify the reports connected to COVID-19. Compared to UAVs and VSNs, social sensing has a broader outreach, but suffers from inconsistent reliability. On the other hand, UAVs and VSNs are fitted with arrays of sensors (e.g., temperature, humidity, and air quality sensors, cameras, microphones)~\cite{erdelj2017help} that allow them to sense COVID-19 related events with substantial fidelity~\cite{rashid2019collabdrone}. However, they are limited in sensing scope and possess partial autonomy~\cite{rashid2019sead}. Leveraging the collective strengths of UAVs and VSNs with social sensing can potentially accelerate the discovery of COVID-19 related events. The reliable and high quality measurements provided by physical sensors naturally complement the uncertain estimation and broader sensing scope of social sensing. Driven by the social signals, the mobility and agility of UAVs and VSNs can allow them to be quickly sent to COVID-19 prone areas or hot zones to collect real-time evidence (e.g., people loitering on streets or gathering in larger groups) and ascertain whether the reported cases actually exists before sending out medical teams or law enforcement~\cite{erdelj2017help}. 

A few possible courses of work can focus on either integrating social sensing with UAVs, namely social drone~\cite{rashid2020socialdrone}, or with VSNs, namely social car~\cite{rashid2019socialcar} to sense the neighborhood of COVID-19 affected areas for unwanted crowds, open pharmacies or emergency supply stores, and so on. Social drone based approaches can be further integrated with computational modeling (e.g., disease propagation models) to enhance the COVID-19 detection process~\cite{rashid2020compdrone}. A set of open research questions in these applications are: i) how to leverage the noisy social signals to quickly guide drones and cars to locations of interest? ii) How to accommodate various constraints imposed by the physical world (e.g., deadlines of urgent cases like dying patients, limited availability drones and their limited flight times)? iii) How to leverage the observations collected by the drones (e.g., unwanted crowds) to improve the social sensing process? Probable solutions that holistically solve the above challenges in the context of CovidSens systems are yet to be developed. 
	\section{Conclusion}\label{sec:conclusion}
In this paper, we introduce CovidSens, a new vision of reliable social sensing based information distillation and risk alerting systems to monitor the COVID-19 spread and study the transmission dynamics of the contagious disease. We highlight a few key challenges in CovidSens applications including data collection, reliability, scalability, modality, presentation, and misinformation spread. By harnessing interdisciplinary techniques, CovidSens can combine the collective strengths of social sensing with AI as well as human intelligence to perform real-time analyses on the obtained epidemiological data. CovidSens can yield more timely and accurate prediction of the COVID-19 spread which may subsequently be presented to end users through a collection of rich mobile apps and UAVs. We hope this paper will uphold CovidSens as an important avenue for guiding research to tackle the current COVID-19 pandemic around the world. 
	
\section*{Acknowledgment}

This research is supported in part by the National Science Foundation under Grant No. CNS-1845639, CNS-1831669, Army Research Office under Grant W911NF-17-1-0409. The views and conclusions contained in this document are those of the authors and should not be interpreted as representing the official policies, either expressed or implied, of the Army Research Office or the U.S. Government. The U.S. Government is authorized to reproduce and distribute reprints for Government purposes notwithstanding any copyright notation here on.

\bibliographystyle{spmpsci}      
\bibliography{refs}   

\end{document}